\documentstyle[aps,multicol,epsf]{revtex}
\def\dir{.}
\def\gaeq{\ \lower-1.2pt\vbox{\hbox{\rlap{$>$}\lower5pt\vbox{\hbox{$\sim$}}}}\ }
\def\laeq{\ \lower-1.2pt\vbox{\hbox{\rlap{$<$}\lower5pt\vbox{\hbox{$\sim$}}}}\ }
\begin{document}
\title{Conformations of Randomly Linked Polymers}
\author{Yacov Kantor}
\address{School of Physics and Astronomy, Tel Aviv University, 
Tel Aviv 69 978, Israel}
\author{Mehran Kardar}
\address{Department of Physics, Massachusetts Institute of  
Technology, Cambridge, MA 02139, U.S.A.}
\date{\today}
\maketitle
\begin{abstract}
We consider polymers in which $M$ randomly selected 
pairs of monomers are restricted to be in contact. Analytical 
arguments and numerical simulations show that an {\it ideal} (Gaussian)
chain of $N$ monomers remains {\it expanded} as long as $M\ll N$; 
its mean squared end to end distance growing as $r^2\propto M/N$.
A possible collapse transition (to a region of order unity) is related 
to percolation in a one dimensional model with long--ranged connections.
A directed version of the model is also solved exactly. Based on these 
results, we conjecture that the typical size of a {\it self-avoiding} polymer 
is reduced by the links to $R\gaeq (N/M)^\nu$. The number of links 
needed to collapse a polymer in three dimensions thus scales as
$N^\phi$, with $\phi\gaeq 0.43 $.
\end{abstract}
\begin{multicols}{2}
\narrowtext
\section{Introduction}\label{intro}
Polymers subject to both repulsive and attractive self--interactions 
may have different equilibrium states depending on temperature. 
For example, a homogeneous self--avoiding polymer with short--range 
attractions between its monomers undergoes a collapse 
transition\cite{degennes} from an expanded to a compact state as the 
temperature is lowered through a ``$\theta$--point''. In the expanded 
state, the radius of gyration (root--mean--squared size) $R_g$ of
the polymer scales as $N^\nu$, where $N$ is the number of monomers, 
and $\nu$ (approximately 0.588 in $d=3$ dimensions\cite{kremer}) is the
swelling exponent. The compact state has a finite density, and hence
$\nu=1/3$. Upon collapse, contacts in the polymer (pairs of monomers 
located adjacent to each other in the embedding space) both increase
drastically, and change qualitatively. The description of such changes
constitutes another important characteristic of the collapse transition.

Statistics of such contacts is even more important in the characterization 
of the equilibrium states of {\it heterogeneous polymers}\cite{shakhnov} 
such as biomolecules.  Dynamics of protein folding is also influenced
by contacts between aminoacids (see, e.g.\cite{Camacho}).
While in problems of this type the contacts between 
monomers are {\it temporarily} generated in the process of thermodynamic 
equilibrium, it is interesting to ask the inverse question of whether the 
configuration of a polymer can be described by specifying {\it permanent} 
contacts between its monomers. The issue of permanent contacts has 
also been extensively addressed in the context of rubber elasticity and the 
vulcanization process\cite{Edwards,Goldbart,LiT,Panykov}, where the typical situation is a 
polymer melt with permanent crosslinks between the polymers.

Gutin and Shakhnovich (GS)\cite{GS} analyzed the problem of a single 
polymer chain in which pairs of monomers are forced to remain adjacent
to each other, i.e. are permanently linked. They note that in order to 
establish a meaningful relation between the  distribution of such
contacting pairs, and the behavior of real self--interacting polymers 
without permanent links, the choice of the ensemble of contacts must be
very non--random: The number of random selections of contacting pairs 
of monomers is of the order of $N^N$, while the number of spatial 
configurations increases only exponentially with $N$. It is thus not
possible to find a simple correspondence between the two random
ensembles. GS suggest that a correlated selection of constrains
is needed to generate appropriate spatial configurations. 

Recently, Bryngelson and Thirumalai (BT)\cite{BT} considered
a related problem in which links are {\it randomly} established between 
pairs of monomers on an ideal (i.e. non--self--interacting) polymer. 
The unconstrained ideal chain is expanded with 
$R_g\propto N^{1/2}$. By comparing variational estimates to the free
energies of expanded and collapsed states, BT argue that increasing
the number of (uncorrelated) links causes the polymer to
collapse into a state in which $R_g$ is independent of $N$. 
In particular, they conclude that for a generic set of constraints, 
in which the typical distance $\ell$ (measured along the backbone)
between linked monomers is of the order $N$, it suffices to have a 
negligible density of the constrains ($\sim 1/\ln N$) for such a 
collapse to occur. 

In this work we consider several models of polymers with
randomly linked monomers. An essential feature of all these 
models is that the links along the polymer are selected in
{\it uncorrelated} fashion. Unlike the previous work of BT,
which concentrated on estimates of the free energy, we
directly measure the spatial extent of the polymer. 
For ideal chains, we derive exact lower bounds which prove 
that uncorrelated links cannot 
cause the polymer to collapse. This is confirmed by extensive
numerical simulations. Based on these results, we conjecture
that, quite generally, the presence of $M$ random links 
reduces the typical size of a swollen polymer to $R\gaeq
(N/M)^\nu$. For ideal chains, a collapse (to $R\sim O(1)$)
occurs only when the number of links is of the order of $N$.
However, it should be easier to collapse self-avoiding chains
to a compact globular state with $R\propto N^{1/3}$.

The remainder of the paper is organized as follows. The simplest
model of an {\it ideal chain} with permanent links is introduced in
Sec.~\ref{model}. We show that calculation of the squared
end to end distance is equivalent to determining the resistance
of a related resistor network. This equivalence provides a
powerful numerical tool that is exploited in the following sections.
Section \ref{near} focuses on the case where the contacts are
formed only between nearby monomers. We demonstrate that 
even a finite density of such contacts does not lead to a
collapsed state. Analogies to percolation in one dimension 
suggest that collapse requires contacts between far away points.
Accordingly, in Sec.~\ref{far}, we consider links between any two 
randomly chosen points on the chain. In this limit, a rigorous lower 
bound indicates that a finite contact density is necessary to cause 
a collapse. Several features of these models
are shared by a simple directed polymer with randomly forced
links to the origin. As discussed in Sec.~\ref{direct}, this model
has the advantage of being exactly solvable. Further discussions
of the results, and their possible extension to {\it self-avoiding}
polymers, appear in conclusion (Sec.~\ref{disc}).

\section{Model}\label{model}
\begin{figure}
\epsfysize=2.0truein
\vbox{\vskip 0.15truein \hskip 0.4truein
\epsffile{\dir/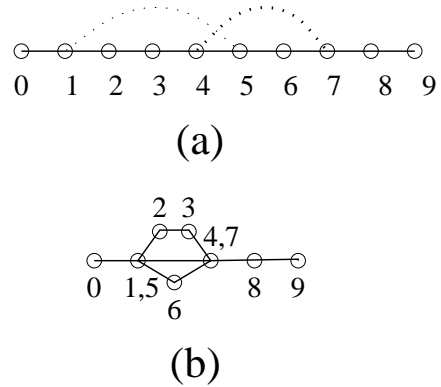}
\vskip 0.15truein
\caption{(a) A schematic drawing of a chain with two linked pairs;
(1,5) and (4,7). The circles represent monomers, the solid lines depict 
bonds between neighboring monomers, and the dashed lines connect
pairs of monomers which are forced to remain in contact.
(b) The same chain as in (a), but with every pair of linked monomers
depicted by a single circle.}
\label{FigA}
}
\end{figure}

Following BT, we use a discrete 
Gaussian chain of $N+1$ monomers, subject to a Hamiltonian
\begin{equation}\label{betaH}
\beta H= {d\kappa\over2}\sum_{i=1}^N(\vec{r}_i-\vec{r}_{i-1})^2\ ,
\end{equation}
to describe the ideal polymer.
Here $\vec{r}_i$ is the position of the $i$th monomer in the $d$
dimensional embedding space, $\beta=1/(k_BT)$, and $\kappa$ 
is the inverse of the mean--squared distance between adjacent
monomers (Kuhn length). We next select $M$ pairs of monomers
$\left\{ k_j^{(1)},k_j^{(2)}\right\}$, for $j=1,2,\cdots,M$ and constrain
each pair to remain in contact. The statistical weight of the 
configurations is now given by
\begin{equation}\label{weight}
P\left[\{\vec{r}_i\}\right]\propto\exp^{-\beta H}
\prod_{j=1}^M\delta^{(d)}
\left(\vec{r}_{k_j^{(1)}}-\vec{r}_{k_j^{(2)}}\right).
\end{equation}
Fig.~\ref{FigA}a depicts schematically a simple case of such a
polymer, with dashed lines connecting the linked monomers. The
same chain is redrawn in  Fig.~\ref{FigA}b in such a way that the 
paired  monomers are placed at the same point in space.

The spatial extent of the polymer can be characterized by the thermal 
average (with weight $P$) of the squared end to end distance 
$r^2\equiv \langle\left(\vec{r}_0-\vec{r}_N\right)^2\rangle$,
or by the squared radius of gyration $R_g^2$ which is the average of
$r_{lm}^2\equiv\langle\left(\vec{r}_l-\vec{r}_m\right)^2\rangle$
over all pairs of monomers $\{l,m\}$. We note that the fluctuations 
of the polymer in each of the $d$ perpendicular space directions is 
independent of the others. Thus $r^2$ is simply the sum of $d$ 
contributions which can be calculated separately by considering 
one dimensional chains. Therefore, without loss of generality we 
restrict ourselves to a {\em one dimensional} Gaussian chain.

It is convenient to recast the problem in terms of a resistor network: 
Mathematically, calculation of $r^2$ for a Gaussian polymer with 
constraints is identical to finding the resistance of a chain built of 
elementary resistors with conductances $\kappa$, in which the 
pairs of sites $\left\{ k_j^{(1)},k_j^{(2)}\right\}$ are connected by zero
resistance links (shorts). Thus, Fig.~\ref{FigA} may also represent
an electrical circuit, where the dashed lines indicate the shorts. 
This {\em exact} correspondence\cite{SCK} holds for any arbitrary
Gaussian network: The dissipation on a link $ab$ is given by 
$\kappa_{ab}(\phi_a-\phi_b)^2$, where $\kappa_{ab}$ is the 
link conductance and $\phi_a$ is the potential on a node. This is 
analogous to the appropriate elastic term in the Gaussian Hamiltonian.
The resistance between any two nodes $j$ and $k$ is calculated by 
minimizing the overall dissipation under applied external voltage. 
For  a quadratic form this minimization is equivalent to calculating 
averages of $\left( \phi_j-\phi_k \right)^2$ with the Gaussian weight 
built using this form. Therefore, our task is reduced to calculating 
the resistance of networks such as the one in Fig.~\ref{FigA}b. 
In the remainder of the paper we shall use the terminology of the
Gaussian polymer and of the resistor network interchangeably.

Note that $\kappa$, which is the force constant for the Gaussian 
chain or the conductance of an elementary resistor, appears 
only as a overall prefactor ($1/\kappa$) in calculations of $r^2$ 
or resistance. Thus without loss of generality we set $\kappa=1$, 
making these quantities dimensionless. Calculating the resistance
of a chain with a specified set of links is now accomplished using 
elementary methods: The configuration is first recast in the form
of a simple electrical network, as in the process leading from 
Fig.~\ref{FigA}a to Fig.~\ref{FigA}b. At this point each resistor is
assigned a unit resistance. Pairs of resistors which are in series 
or parallel are replaced by effective resistors. Repeated 
application of this process leads to a network of not more than 
$M$ nodes connected by effective resistors. (For low densities
of shorts, the number of nodes is much less than $M$. For example,
reduction by series and resistor rules is sufficient to completely
eliminate all internal nodes in Fig.~\ref{FigA}b.) 
Finally, the resistance of the reduced network is calculated
by solving a system of linear equations. The number of unknowns
(and equations) is of order of (or smaller than) $M$, and thus
much smaller than $N$. Therefore, for each $N$ and $M$, we 
could easily average our numerical results over large numbers
(up to 1600) of configurations with randomly distributed links.

\section{Nearby links}\label{near}
We start by considering links that join monomers that are close-by
along the chain. As a simple example consider a very long chain 
($N\gg1$) with $M$ contact pairs scattered randomly along the
chain: The position $k_j^{(1)}$ of the first monomer of the $j$th pair
is chosen with uniform probability anywhere along the chain, while 
the second member of the pair is located at $k_j^{(2)}=k_j^{(1)}+\ell$, 
i.e. the distances between any two members of a pair (measured 
along the backbone) are fixed at $\ell$. (We assume that $\ell\ll N$.) 

For the corresponding electrical circuit, it is obvious that the total
resistance is proportional to $N$ (because the shorts are local).
The problem is characterized by the density $n=M/N$, and 
``coverage" $c\equiv n\ell$ of the links. As long as the coverage is
small ($c\ll 1$), the different pairs do not bridge over overlapping
strands of the chain; the total resistance is obtained simply by 
removing the part of the chain that is shortened as 
$r^2=N-M\ell=(1-c)N\approx{\rm e}^{-c}N$. 

When $c$ is comparable or larger than unity, the resistance of the 
chain drops significantly. 
However, it can still be bounded from below by the total resistance 
of the resistors which are not bridged by the shorts.
An ``unbridged resistor'' is such that there are no shorts which begin 
to its left and end to its right. Since the probability of such a condition
for each resistor is $(1-n)^\ell$, a lower bound on the resistance is 
given by $(1-n)^\ell N$. The continuum limit (where the discreteness
of the chain can be disregarded) is reached when $n\ll1$ and 
$1\ll\ell\ll N$. In this limit the bound becomes ${\rm e}^{-c}N$. 
(Note that $c$ does not have to be small.) This lower bound shows 
that the chain is not collapsed for any coverage, i.e. $r^2\sim N$, 
although the prefactor may be very small when the links are
dense.

\begin{figure}
\epsfysize=3.0truein
\vbox{\vskip 0.15truein \hskip 0.4truein
\epsffile{\dir/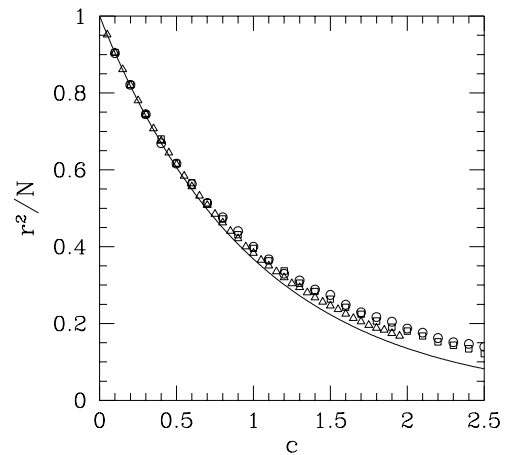}
\vskip 0.15truein
\caption{Normalized resistance, or $r^2$, of a chain of $N=2560$ 
monomers, for links of fixed lengths $\ell=8$ (triangles), $\ell=32$ 
(squares), and $\ell=128$ (circles) as a function of the coverage 
$c=M\ell/N$.}
\label{FigB}
}
\end{figure}

The results of Monte Carlo simulations on this model are depicted in
Fig.~\ref{FigB}. In the continuum limit, the resistance is expected to
have the form $g(c)N$, where $g$ is some unknown function.
The collapse of the data for different values of $\ell=8$, 32, and 128
confirms this expectation. There are slight systematic deviations
for $\ell=8$ which are due to the discreteness of the chain. The
lower bound of $e^{-c}$ is indicated by the solid line in this figure,
and is quite a good estimate for small values of coverage.

There are close analogies between random resistor networks and
percolation. In particular, the collapse of the Gaussian chain can be 
related the presence of infinite cluster of shorts. In the corresponding
percolation problem, long--range shorts are added to a one--dimensional
system\cite{grimmet} in uncorrelated fashion. The results of this section 
were  obtained for the simple case where the shorts connect
points at a fixed distance $\ell$.  However, the conclusions can be 
generalized to the case when $\ell$ is  randomly distributed with a
probability $p(\ell)$.

An important characteristic of long--range percolation is the coverage
$c\equiv \sum_{\ell=1}^\infty \ell p(\ell)$. As long as $c$ is finite, the
shorts do not form an infinite cluster\cite{grimmet}. The resistance 
(or $r^2$) is then proportional to $N$, with a lower bound of 
${\rm e}^{-c}N$. Power--law distributions $p(\ell)\approx B/\ell^\alpha$ 
(for $\ell\to\infty$) are frequently used to describe long--range percolation. 
For  $\alpha>2$ there is a finite $c$, leading to the situation described above.
For $\alpha<1$, the sum $n\equiv\sum_{\ell=1}^\infty p(\ell)$ diverges,
i.e. the number of contacts per monomer is infinite. In this case, an
infinite clusters of shorts always exists, although the situation does
not correspond to a realistic physical model, since $M/N\to\infty$.
A somewhat more realistic situation occurs for $1<\alpha<2$,
where it can be shown\cite{schulman} that for $n<{1/ 2}$ no 
infinite cluster is formed. For $n>{1/ 2}$, depending on the details
of $p(\ell)$ (e.g., the value of the constant B), percolation may or 
may not occur\cite{newman}. This again corresponds to $M\approx N$.

\section{Distant Links}\label{far}
The distributions with $\alpha<2$ cannot be directly used to describe 
the behavior of a {\em finite} polymer, because the divergence of
$c$ implies the presence of strong finite size effects. We shall, 
therefore, consider the extreme case of a broad distribution of
$\ell$  by assuming that for finite $N$ the typical $\ell$ is of order 
of $N$. The simplest possible situation is obtained when $k_j^{(1)}$ 
and $k_j^{(2)}$ of the $j$th link are selected independently and
uniformly among all monomers. We are interested in calculating
$r^2$ for such a chain with $M$ links. 

We immediately notice a simple scaling argument: Consider a chain 
of length $N$ with $M$ links located at $\left\{ k_j^{(1)},k_j^{(2)}\right\}$, and 
compare it with another chain of $\lambda N$ monomers with links
at $\left\{ \lambda k_j^{(1)},\lambda k_j^{(2)}\right\}$. Clearly, $r^2$ of the 
latter sequence is exactly $\lambda$ times larger than the former.
(We consider the limit $N\gg 1$ where the discreteness effects
can be disregarded.) The corresponding probabilities of finding 
such randomly linked chains are identical. (More precisely, the 
probability for a link in the first ensemble to be located between 
 $x$ and $x+dx$ is equal to the probability for a link in the second 
ensemble to be located between $\lambda x$ and $\lambda(x+dx)$.) 
From the relation $r^2_{M,\lambda N}=\lambda r^2_{M,N}$, it 
immediately follows that $r^2=f(M)N$. To obtain a collapsed state 
of the chain we must have $f(M)\sim 1/N$. We shall show that such 
a small value of $f$ is reached only when $N\sim M$, when 
essentially every monomer is paired with another.

We first obtain a lower bound for $r^2$ in this case. The $M$ shorts
break the chain backbone into $2M+1$ segments. The resistance of
the chain is certainly larger than that of the two extremal segments 
at its two ends. In Fig.~\ref{FigA}, this corresponds to distance 
between 0 and 1, plus the distance between 7 and 9. In the limit of large 
$M$, each segment is independently taken from an exponential
probability distribution (see Sec.~\ref{direct}) with mean size
$N/(2M+1)$. The resistances of the two end
segments thus add up to $2N/(2M+1)\approx N/M$. Therefore, for 
large $M$ we have $f(M)\equiv r^2/N>1/M$. Note, that this bound ensures
the absence of a collapsed state for vanishing density $M/N$,
contradicting a prediction of Ref.~\cite{BT}. 

We were unable to derive a satisfactory upper bound for $f(M)$.  
It can be crudely argued that it is bounded from above by $(\ln M)/M$: The 
resistance of the chain should be smaller than that of a single path going 
from one end of the chain to the other, either by the way of links or through
the shorts. Since (for small $M$) the number of different $L$--step paths 
(on the space of $2M+1$ segments) beginning from monomer 0,
increases exponentially with $L$, the probability that at least one of the 
paths reaches the point $N$ becomes of the order of unity when $L\sim\ln M$. 
The resistance of the path which has reached the end--point can now 
be estimated by multiplying the resistance of  a typical segment,
approximately $N/(2M)$, by the number of segments, $\sim\ln M$. 
Therefore, this overestimate of the resistance is approximately $N\ln M/M$.

\begin{figure}
\epsfysize=3.0truein
\vbox{\vskip 0.15truein \hskip 0.4truein
\epsffile{\dir/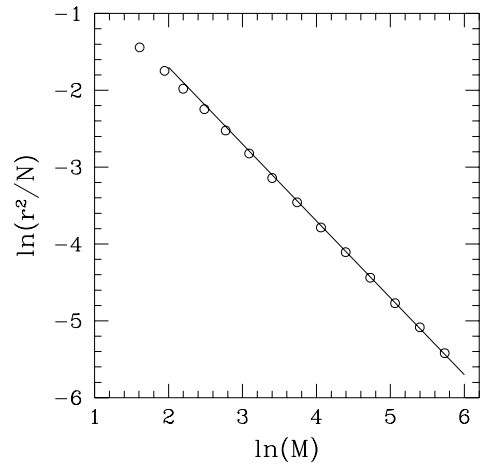}
\vskip 0.15truein
\caption{Logarithmic plot of the scaled resistance $r^2/N$, as a 
function of number of links $M$. Each point represents an average 
over 1600 randomly linked chains. The slope of the solid line is $-1$.}
\label{FigC}
}
\end{figure}

We confirmed numerically the scaling form of $r^2$ for several values
of $N$. Fig.~\ref{FigC} depicts (on a logarithmic scale) the scaled 
resistance $f=r^2/N$, as a function of $M$, for a single value of 
$N=2560$. Every point on this figure represents an average  over 
1600 configurations of random links. The numerical results 
gradually converge to a slope of $-1$, as depicted by the solid line.
A least squares fit to {\em all} points of the figure produces a 
slope of 0.97, and the curve cannot be fitted as $\ln M/M$. In fact,
we conclude that $f(M)\approx 1.5/M$; with a prefactor that is 
surprisingly close to the value of 1 which appears in our simple 
lower bound.

\section{Directed Polymers}\label{direct}
In this section we introduce a simpler model in which a random set
of monomers is linked to the origin. Alternatively, the model describes
a {\it directed polymer} in $d+1$ dimensions, in which certain certain
points are linked to a line at the origin. In the latter example, $\vec{r}_i$ 
denotes the transverse coordinate of the $i$th monomer. The $M$
randomly chosen monomers $\{ k_j\}$, for $j=1,2,\cdots,M$ are linked
to the origin, resulting in the statistical weight,
\begin{equation}\label{dpweight}
P\left[\{\vec{r}_i\}\right]\propto\exp^{-\beta H}
\prod_{j=1}^M\delta^{(d)}
\left(\vec{r}_{k_j}\right).
\end{equation}

The constraints break the chain into $M+1$ independent segments
of lengths $ \left\{ s_\alpha \right\}$, with $\alpha=1,2,\cdots,M+1$.
This model is easily solvable in the limit of large $M$, where
it is sufficient to apply the constraint $\sum_\alpha s_\alpha=N$ 
{\it only on average}. Subject to this constraint, the joint probability of 
segment lengths is maximized for a product of independent exponential 
distributions
\begin{equation}\label{Poisson}
p(s)={M+1\over N}\exp\left( -{M+1\over N}s \right).
\end{equation}
For each segment, we thus have $\left\langle s^m\right\rangle\approx
m! (N/M)^m$.

The end to end distance in this case is given simply by the contribution
of the two end segments, and $\left\langle r^2\right\rangle=2N/M$. 
We can also calculate an average of the squared distance from the origin,
\begin{equation}\label{Rg}
R^2\equiv{1\over N}\sum_{i=0}^N r_i^2={1\over N}
\sum_{\alpha=1}^{M+1} r(s_\alpha)^2,
\end{equation}
where we have taken advantage of the independence of segments.
It is easy to show that each segment contributes $r(s)^2=s^2/6$, 
resulting in
\begin{equation}\label{Rgave}
\left\langle R^2\right\rangle={1\over 3}\left( {N\over M} \right).
\end{equation}
(We note that $R^2$ does not coincide with $R_g^2$, but differs from
it only by a term of order $N/M^2$, which becomes negligible for large $M$.)
We can also consider cases where the chains are non-ideal, such as
directed polymers in a random medium\cite{DPRM}. In such
cases, each segment wanders away from the origin by an amount
$s^\nu$ with $\nu>1/2$. The overall radii of the randomly linked 
polymer are then characterized by the scale $(N/M)^\nu$.

\section{Discussions}\label{disc}
Most of the results described so far apply to ideal chains. However,
we may argue that some of conclusions are expected to hold for
self-avoiding polymers. Consider, for example, the lower bound 
obtained in Sec.~\ref{far}. Even for a self-avoiding walk, it is 
reasonable to expect that the overall polymer size is larger than the
contributions of its two end segments. Although the interactions
amongst the segments make an exact argument difficult, it is
probably reasonable to assume that each end segment has a
characteristic size proportional to $s^\nu$ with $s\approx N/M$,
where $\nu$ is the swelling exponent. The numerical results of
Sec.~\ref{far}, and the analytical forms of Sec.~\ref{direct}, suggest
that this bound is generally satisfied. We thus conjecture that the
typical size of a self-avoiding chain with $M$ random links is 
given by
\begin{equation}\label{conjecture}
\left\langle R\right\rangle\propto \left( {N\over M }\right)^\nu .
\end{equation}

Assuming the validity of the above conjecture, we can ask how
many links are necessary to compactify a chain. An ideal chain
in its compact state is localized to a region of size $R\sim O(1)$.
This is achieved only with an extensive number of links 
$M\propto N$. For a self-avoiding chain, however, the compact
state has finite density, and hence $R\propto N^{1/d}$. Comparing
with Eq.~(\ref{conjecture}), suggests that such compactification
is achieved if the number of links scales as
\begin{equation}\label{phi}
M\propto N^\phi\quad{\rm with}\quad \phi=1-{1\over d\nu} .
\end{equation}
For a self-avoiding polymer in $d=3$, $\phi\approx 0.43$, and 
$\phi=1/3$ in $d=2$. (While this certainly gives the minimum
number of bonds necessary for collapse, its sufficiency remains
to be established.)

The above result suggests that it is much easier to compactify a
self-avoiding polymer. However, it says very little about the
final structure of the compact state. From the perspective of
protein folding studies\cite{GS}, the resulting state is most likely 
a compact globule; a liquid-like state with extensive entropy.
Additional links would then be needed to freeze this compact object
into a unique configuration. The radius of gyration is then not
a good discriminator of the state of the macromolecule.

At the completion of this work, we became aware of a recent paper by 
Solf and Vilgis\cite{SV}.  Their starting point is a randomly crosslinked
Gaussian network, motivated by the Deam and Edwards \cite{Edwards}
model of polymeric gels.  Although they consider more general networks, 
they also perform simulations on the model of Sec.\ref{far}. They measure
the radius of gyration, finding $R_g^2\approx 0.26 N/M$, consistent
with our results. Interestingly, the ratio of  $R_g^2$ to $r^2$ (calculated
in our paper) is approximately 1/6, as in the case of an unconstrained ideal chain.

\acknowledgments 
This work was supported by the US--Israel BSF grant 
No. 92--00026, by the NSF through grant No. DMR--94--00334, and 
the PYI program (MK).

\end{multicols}
\end{document}